\documentclass[12pt]{article}
\usepackage{graphicx}
\usepackage{amsfonts}
\usepackage{amssymb,amsmath}
\usepackage{latexsym}
\usepackage{color}
\usepackage{cite}
\input{colordvi.tex}

\begin{document}

\begin{titlepage}

\begin{center}

{\Large \bf  
Density Fluctuations in Thermal Inflation \\ 
and Non-Gaussianity
}

\vskip .45in

{\large
Masahiro Kawasaki$^{1,2}$, 
Tomo Takahashi$^3$ and 
Shuichiro Yokoyama$^4$
}

\vskip .45in

{\em
$^1$Institute for Cosmic Ray Research, University of Tokyo,
Kashiwa 277-8582, Japan  \vspace{0.2cm} \\
$^2$Institute for the Physics and Mathematics of the Universe,
University of Tokyo, Kashiwa, Chiba, 277-8568, Japan \vspace{0.2cm}\\
$^3$Department of Physics, Saga University, Saga 840-8502, Japan \vspace{0.2cm}\\
$^4$Department of Physics and Astrophysics, Nagoya University,
Aichi 464-8602, Japan
}

\end{center}

\vskip .4in

\begin{abstract}
  We consider primordial fluctuations in thermal inflation
  scenario.  Since the thermal inflation drives about 10 $e$-folds
  after the standard inflation, the time of horizon-exit during
  inflation corresponding to the present observational scale shifts
  toward the end of inflation. It generally makes the primordial power
  spectrum more deviated from a scale-invariant one and hence renders
  some models inconsistent with observations. We present a mechanism
  of generating the primordial curvature perturbation at the end of
  thermal inflation utilizing a fluctuating coupling of a flaton field
  with the fields in thermal bath.  We show that, by adopting the
  mechanism, some inflation models can be liberated even in the
  presence of the thermal inflation.  We also discuss non-Gaussianity
  in the mechanism and show that large non-Gaussianity can be
  generated in this scenario.
	
\end{abstract}

\end{titlepage}

\setcounter{page}{1}

\section{Introduction}

The thermal
inflation~\cite{Yamamoto:1985rd,Binetruy:1986ss,Lazarides:1992gg,Lyth:1995hj,Lyth:1995ka,Barreiro:1996dx,Asaka:1999xd},
a mini-inflation which occurs long after the standard inflation, has
been discussed, in particular, as a possible solution to the
cosmological moduli problem
\cite{Coughlan:1983ci,Banks:1993en,deCarlos:1993jw}, in which moduli
particles would lead to various cosmological difficulties such as
destroying light element synthesized by BBN, too much contribution to
X($\gamma$)-ray background radiation, overclosure of the universe and
so on.  The thermal inflation can produce very large entropy to dilute
the moduli density sufficiently and thus can be a solution for the
moduli problem.

Here we would like to discuss another aspect of the thermal inflation.
When one considers the thermal inflation, primordial fluctuations are
usually assumed to be generated from the primordial inflation.
However notice that the thermal inflation drives the number of
$e$-folds of about $10$ due to the large entropy production.  This
means that fluctuations corresponding to the present observational
scales exit the horizon at later time compared with the case of no thermal
inflation.  Since the time of horizon-exit becomes closer to the end
of inflation, the primordial power spectrum is rendered to be more
tilted in most cases. Hence even an inflation model consistent with
observations in the absence of the thermal inflation can be pushed to
outside the allowed range.  For example, here let us consider the
quadratic chaotic inflation model, where the inflaton potential is given by
 $V \propto \phi^2$. In the
usual case, the curvature perturbations probed by cosmological
observations are assumed to correspond to the fluctuations which exit
the horizon when the number of $e$-folds is $ N_{\rm inf} = 50-60$.
By adopting this number, the spectral index is $n_s \sim 0.96$ and the
tensor-to-scalar ratio is $r \sim 0.13$, which is consistent with
current observations such as WMAP5 \cite{Komatsu:2008hk}.  However,
when the thermal inflation occurs after the standard primordial inflation, the
number of $e$-folds reduces by about 10, and then the spectral index is
more red-tilted and the tensor-to-scalar ratio is more increased. When
$N_{\rm inf}=40$, the tensor-to-scalar ratio is $r\sim 0.2$, which is
inconsistent with current observations and invalidates the model.
Thus in this respect, the thermal inflation can also affect the
predictions of primordial fluctuations.

In fact, the primordial fluctuations are not necessarily generated
from the inflaton fluctuations.  Another light scalar field such as
the curvaton \cite{Enqvist:2001zp,Lyth:2001nq,Moroi:2001ct}, modulus
in the modulated reheating scenario \cite{Dvali:2003em,Kofman:2003nx}
and so on are also known to generate (almost) scale-invariant and
adiabatic primordial fluctuations consistent with
observations\footnote{
  Other mechanisms have also been discussed in the literatures
  \cite{Kolb:2004jm,Lyth:2005qk,Alabidi:2006wa,Sasaki:2008uc}.
}.  Although by adopting these scenarios, one can alleviate the above
mentioned issue, however, here we propose another mechanism which can
naturally arise in the framework of the thermal inflation.  During the
thermal inflation, the effective potential of a flaton, a flat
direction in supersymmetric models, is lifted up by a thermal effect
due to the coupling between the flaton and particles in thermal bath.
The end of the thermal inflation is controlled by the strength of the
thermal effect, or the coupling. If the coupling depends on 
some other scalar field and this scalar field fluctuates, the
end of the thermal inflation also differ from place to place in the
Universe via the fluctuations of the coupling.  If the scalar field is
light enough during inflation, almost scale-invariant density
fluctuations can be achieved similarly to the case of the modulated
reheating\footnote{
  In Ref.~\cite{Matsuda:2009yt} the author has presented a mechanism
  of generating large scale curvature perturbation through the
  inhomogeneous cosmological phase transitions in the early universe.
  The mechanism proposed here is similar to this kind.
}.

Another important issue in the primordial fluctuations is the Gaussian
nature of the fluctuations.  In fact, although recent observations are
consistent with purely Gaussian fluctuations, but they may suggest
that primordial local type non-Gaussianity\footnote{
  Two types of non-Gaussianity are often discussed
  in the literatures, one is the so-called local type and the other is
  the equilateral type.  In this paper, we only focus on the local
  type non-Gaussianity.
} is large compared to that of the standard (single-field) inflation
model \cite{Komatsu:2008hk,Smith:2009jr}.  Since some of the
alternative mechanisms such as the curvaton and the modulated
reheating scenarios can also produce large non-Gaussianity, they have
been the subject of intense study \cite{nonGcurvaton,nonGmodreheat}.
We show that large non-Gaussianity can be generated in the mechanism
as well. Thus the mechanism proposed here would be interesting in this
respect too.

The structure of this paper is as follows. In the next section, we
briefly describe the thermal inflation model which we consider here.
Then in Section~\ref{sec:deltaN}, we summarize the formalism to
discuss the curvature perturbation in the model and some observational
quantities such as power spectrum, bispectrum and trispectrum 
or nonlinearity parameters.  In Section~\ref{sec:nonG}, we present a mechanism of
generating the curvature perturbation at the end of thermal inflation
and also discuss its non-Gaussianity.  In Section~\ref{sec:model}, we
discuss the current observational constraints for several simple
inflation models in our scenario.  The final section is devoted to
summary of this paper.
 
\section{Basic Picture of thermal inflation} \label{sec:thermal}

In this section, we briefly review the idea of the thermal
inflation~\cite{Lyth:1995ka}.  The thermal inflation can be realized
by utilizing a flat direction which exists in supersymmetric
theory. Let us call such a direction as a flaton field $\phi$ and
assume that the potential of the flaton field is given by
\begin{eqnarray}
V(\phi) = V_0 - {1 \over 2}m^2 \phi^2 + {\lambda \over 6}{1 \over M_{\rm Pl}^2}\phi^6~,
\label{eq:susypot}
\end{eqnarray}
where the $\phi^2$ term comes from soft supersymmetry breaking and we
have neglected the higher order nonrenormalizable terms
($\phi^{2n+4}$, $n \ge 2$).  The VEV of $\phi$ is given
by $\phi_{\rm vev} \equiv v_{\rm flaton} = \lambda^{-1/4}
m^{1/2}M_{\rm Pl}^{1/2}$ and $V_0 = m^2 v^2_{\rm flaton} / 3$ to have
$V(v_{\rm flaton}) = 0$.  When the flaton field interacts with some
particles in thermal bath, such an interaction gives a thermal
contribution to the potential of the flaton field as
\begin{eqnarray}
V_T = {g \over 2} T^2 \phi^2~,
\end{eqnarray}
where $g$ is an effective coupling between the flaton and particles in
thermal bath and $T$ is the cosmic temperature. Then the total
effective potential of the flaton field in thermal bath is given by
\begin{eqnarray}
V_{\rm eff} = V_0 + {1 \over 2}\left( g T^2 - m^2 \right)\phi^2
 + {\lambda \over 6}{1 \over M_{\rm Pl}^2}\phi^6~.
\end{eqnarray}
After the standard primordial inflation ends followed by the reheating, the
cosmic temperature decreases as the Universe expands. Then, at some
temperature when $V_0$ gets larger than the background radiation
energy density, the Universe is dominated by the false vacuum of the
flaton's potential, which drives a mini-inflation.  When the
temperature of the Universe has dropped down to $T = T_c= m/\sqrt{g}$,
the flaton starts rolling down to the VEV and then the mini-inflation
ends.  The duration of this thermal inflation can be expressed by
using the $e$-folding number as
\begin{eqnarray}
N(t_c, t_{\rm in}) = \int^{t_c}_{t_{\rm in}} H dt
  = 
  \ln \left( {a_c \over a_{\rm in}} \right) = - \ln \left( {T_c \over T_{\rm in}} \right) ~,
\label{eq:thermalefold1}
\end{eqnarray}
where $H$ denotes a Hubble parameter.  Here $T_{\rm in}$ is the
temperature at the onset of the thermal inflation, which is defined as
$\rho_r(T_{\rm in}) = V_0 = m^2 v^2_{\rm flaton} /3$, i.e., $T_{\rm
  in} \sim m^{1/2} v^{1/2}_{\rm flaton}$,
where  $\rho_r$ represents the radiation energy density.
Taking some typical values of $m \simeq
10^2~{\rm GeV}$ and $v_{\rm flaton} \simeq 10^{10}~{\rm GeV}$, we have
$T_{\rm in} \simeq 10^6~{\rm GeV}$ and $T_c \simeq 10^2~{\rm GeV}$ and
hence the size of the $e$-folds during thermal inflation is estimated
as $N(t_c, t_{\rm in}) \simeq 10$, which corresponds to 
the number
$\Delta N_{\rm th}$ to be subtracted from the standard one.  More
precisely $\Delta N_{\rm th}$ is calculated by considering entropy
production due to the thermal inflation.  The total entropy $S$ of the
Universe after thermal inflation increases by a factor $\Delta_S =
s(T_r) a_r^3 /s(T_c)a_c^3$ where $a_r$ and 
$a_c$ are the scale factor at the reheating after the thermal inflation 
and the end of the thermal inflation, respectively. $s$ is the entropy density and 
$T_r$ is the reheating temperature after thermal inflation.
Since the flaton behaves like a matter after the end of the thermal inflation,
the energy density of the flaton just before the reheating is given by
$\rho_{\rm flaton}(T_r) = V_0(a_c/a_r)^3$.
This gives the radiation energy density just after the reheating,
hence the entropy density can be written as
$s(T_r) = 4\rho_r(T_r)/3T_r$.
Then,  
$\Delta_S$ can be 
estimated as
\begin{equation}
  \Delta_S = \frac{4V_0}{3T_r} \frac{45}{2\pi^2g_* T_c^3}
  \simeq 0.01 \frac{v_{\rm flaton}^2}{m T_r},
  \label{eq:entropy}
\end{equation}
where  $g_* (\simeq 100)$ counts the effective
degrees of radiation.  Then $\Delta N_{\rm th}$ is estimated as
\begin{equation}
  \Delta N_{\rm th} = \frac{1}{3}\ln\Delta_S 
  \simeq 12 - \frac{1}{3}\ln\left( \frac{m}{ 10^2~{\rm GeV}}\right)
  - \frac{1}{3}\ln\left( \frac{T_r}{  {\rm GeV}}\right) 
  + \frac{2}{3}\ln\left( \frac{v_{\rm flaton}}{10^{10}~{\rm GeV}}\right).
  \label{eq:efold_inc}
\end{equation}

\section{$\delta N$ formalism and observational quantities} \label{sec:deltaN}

Based on $\delta N$ formalism~\cite{bib:deltaN,Lyth:2004gb}, the curvature
perturbation on the uniform energy density hypersurface, $\zeta$, on
super-horizon scales is given by the difference of the neighbor
background trajectories, which is parametrized by the $e$-folding
number, measured between the initial flat hypersurface and the final
uniform energy density hypersurface.  That is, we have
\begin{eqnarray}
\label{eq:zeta1}
\zeta (t_f) \simeq \delta N (t_f,t_*)~,
\end{eqnarray}
where $N(t_f,t_*)$ denotes the $e$-folding number measured between
those at $t=t_f$ and $t=t_*$.

When the curvature perturbation $\zeta$ originates from fluctuations
of a single scalar field $\sigma$, $\zeta$ at $t=t_f$ is given, up to
the third order, by
\begin{equation}
\label{eq:defs}
\zeta(t_f) =
N_\sigma \delta \sigma_\ast
+ \frac{1}{2} N_{\sigma\sigma} (\delta \sigma_\ast)^2
+ \frac{1}{6} N_{\sigma\sigma\sigma} (\delta \sigma_\ast)^3,
\end{equation}
where $N_\sigma = dN/d\sigma_\ast$ and so on. The power spectrum
$P_\zeta$, the bispectrum $B_\zeta $, and the trispectrum $T_\zeta $
are given by
\begin{equation}
\label{eq:power}
\langle \zeta_{\vec k_1} \zeta_{\vec k_2} \rangle
=
{(2\pi)}^3 P_\zeta (k_1) \delta ({\vec k_1}+{\vec k_2}),
\end{equation}
\begin{eqnarray}
\langle \zeta_{\vec k_1} \zeta_{\vec k_2} \zeta_{\vec k_3} \rangle
&=&
{(2\pi)}^3 B_\zeta (k_1,k_2,k_3) \delta ({\vec k_1}+{\vec k_2}+{\vec k_3}),
\label{eq:bi}
\end{eqnarray}
\begin{eqnarray}
\langle
\zeta_{\vec k_1} \zeta_{\vec k_2} \zeta_{\vec k_3} \zeta_{\vec k_4}
\rangle
&=&
{(2\pi)}^3 T_\zeta (k_1,k_2,k_3,k_4) \delta ({\vec k_1}+{\vec k_2}+{\vec k_3}+{\vec k_4}), 
\label{eq:tri}
\end{eqnarray}
where $B_\zeta$ and $T_\zeta$ can be expressed with the power spectrum
as
\begin{eqnarray}
B_\zeta (k_1,k_2,k_3)
&=&
\frac{6}{5} f_{\rm NL} 
\left( 
P_\zeta (k_1) P_\zeta (k_2) 
+ P_\zeta (k_2) P_\zeta (k_3) 
+ P_\zeta (k_3) P_\zeta (k_1)
\right), \\
\label{eq:def_f_NL}
T_\zeta (k_1,k_2,k_3,k_4)
&=&
\tau_{\rm NL} \left( 
P_\zeta(k_{13}) P_\zeta (k_3) P_\zeta (k_4)+11~{\rm perms.} 
\right) \nonumber \\
&&
+ \frac{54}{25} g_{\rm NL} \left( P_\zeta (k_2) P_\zeta (k_3) P_\zeta (k_4)
+3~{\rm perms.} \right).
\label{eq:def_tau_g_NL} 
\end{eqnarray}
Here $f_{\rm NL}, \tau_{\rm NL}$ and $g_{\rm NL}$ are the
non-linearity parameters often discussed in the literatures.  By
adopting $\delta N$ formalism, these parameters can be calculated as~\cite{Lyth:2005fi}
\begin{eqnarray}
\frac{6}{5}f_{\rm NL}
=
 \frac{N_{\sigma\sigma}}{ N_\sigma^2 }, 
 ~~~~~~~~
\tau_{\rm NL}
=
 \frac{N_{\sigma\sigma}^2}{ N_\sigma^4 },
  ~~~~~~~~
\frac{54}{25} g_{\rm NL}
=
 \frac{N_{\sigma\sigma\sigma}}{ N_\sigma^2 }.
\label{eq:non_linear_param}
\end{eqnarray}
When the curvature perturbation is generated from a single source, the
relation $\tau_{\rm NL} = (36/25)f_{\rm NL}^2$ holds, which is the
case in the following discussion.  Thus when we investigate the
non-linearity of fluctuations, we do not consider $\tau_{\rm NL}$, but
discuss $f_{\rm NL}$ and $g_{\rm NL}$.

\section{Curvature perturbation generated at the end of thermal inflation} 
\label{sec:nonG}

Now we discuss the generation of density fluctuations at the end of
thermal inflation.  We assume that the coupling $g$ depends on a light
scalar field $\sigma$, namely, $g=g(\sigma)$.  Then the coupling $g$
can fluctuate due to the fluctuation of the light field $\sigma$ which
originates to quantum fluctuations during inflation and hence the
fluctuation of the coupling $g$ gives rise to the inhomogeneous end of
thermal inflation.  In such a case, from Eq.~(\ref{eq:thermalefold1})
the curvature perturbation from fluctuations of $\sigma$ at the end of
thermal inflation can be given by, up to the third order,
\begin{eqnarray}
\zeta = \delta N &\! = &\! - {\delta T_c \over T_c} + {1 \over 2} \left( {\delta T_c \over T_c} \right)^2
- {1 \over 3} \left( {\delta T_c \over T_c} \right)^3
\nonumber \\
&\! = \!&{1 \over 2}\left\{
{g' \over g} \delta \sigma_*
+ {1 \over 2}\left[ {g'' \over g} - \left({g' \over g}\right)^2\right]\delta \sigma_*^2
+ {1 \over 6}\left[ 
{g^{\prime\prime\prime} \over g} - 3 {g'' g'\over g^2} 
+ 2 \left({g' \over g}\right)^3 \right] \delta \sigma_*^3  \right \}~,
\end{eqnarray}
where a prime denotes the derivative with respect to $\sigma_\ast$. 
Here we have used $T_c \propto g^{-1/2}$.

Now the power spectrum is written as
\begin{equation}
\mathcal{P}_\zeta = \frac{1}{4}\left( \frac{g'}{g} \right)^2 \left( \frac{H_\ast}{2\pi} \right)^2.
\end{equation}
Since $\sigma$ does not evolve much during inflation, the scale
dependence of the power spectrum comes from the time variation of the
Hubble parameter during inflation. Then the spectral index is given by
\begin{eqnarray}
\label{eq:ns_2}
n_s - 1 \simeq - 2\epsilon~,
\end{eqnarray}
where $\epsilon$ is a slow-roll parameter and defined by 
$\epsilon = (M_{\rm pl}^2/2) (V_\phi / V)^2$.
Here $V$ is the potential for the inflaton~$\phi$ and $V_\phi = d V /d\phi_\ast$. Notice that in the standard
inflation scenario, the spectral index is given by $n_s -1 = -6
\epsilon + 2 \eta$ with $\eta = M_{\rm pl}^2(V_{\phi\phi} /V)$ being another slow-roll parameter.

To discuss non-Gaussianity of the curvature perturbation in the
mechanism, we give the non-linearity parameters defined in the
previous section.  For the bispectrum, one usually uses the parameter
$f_{\rm NL}$ which is given by
\begin{eqnarray}
{6 \over 5}f_{\rm NL} = 2 
\left[ { g'' / g \over \left( g' / g \right)^2} - 1\right]~.
\end{eqnarray}
For the trispectrum, the non-linearity parameter $g_{\rm NL}$ is
\begin{eqnarray}
{54 \over 25} g_{\rm NL} = 4
\left[ 
{ g^{'''} / g \over \left( g' / g \right)^3} - 3 {g^{''} / g \over \left( g' / g \right)^2} + 
2 \right]~.
\end{eqnarray}

To investigate observational quantities in more detail, we assume a
simple functional form of $g$ as
\begin{eqnarray}
\label{eq:form_g}
g = g_0 \left( 1 + {1 \over 2} {\sigma^2 \over M^2}\right)~,
\end{eqnarray}
with $g_0$ and $M$ being a coupling constant and some scale,
respectively.  With this form of $g$, the curvature perturbation
generated from the fluctuations of $\sigma$ is given by
\begin{eqnarray}
&& \zeta_\sigma \simeq 
\left( {\sigma_* \over M} \right)^{2} {\delta \sigma_* \over \sigma_*}.
 \label{eq:curvamp}
\end{eqnarray}
In general, however, fluctuations from the inflaton can also be
generated. The curvature perturbation $\zeta_{\rm inf}$ from the inflaton $\phi$
can be expressed as
\begin{eqnarray}
\zeta_{\rm inf} \simeq { 1 \over \sqrt{ 2\epsilon}M_{\rm pl}} \delta \phi_*~.
\label{eq:infamp}
\end{eqnarray}
Thus the ratio of the amplitude of the power spectra between those
from the mechanism proposed here and the inflaton is given by
\begin{equation}
\label{eq:ratio}
\frac{P_{\zeta_\sigma}}{P_{\zeta_{\rm inf}}} 
\simeq 
\frac{1}{2} \epsilon \left( {\sigma_\ast \over M }\right)^2 
\left( { M_{\rm pl} \over M }\right)^2. 
\end{equation}
On the other hand, if we assume that $ (\sigma_* / M)^2 \ll 1$, the
non-linearity parameter $f_{\rm NL}$ can be written as
\begin{eqnarray}
&&f_{\rm NL} \simeq \left( {\sigma_* \over M} \right)^{-2}~.
\label{eq:nonlinearity}
\end{eqnarray}
The constraint on the non-linearity parameter from the WMAP5~\cite{Komatsu:2008hk,Smith:2009jr} is $f_{\rm NL} <
\mathcal{O}(100)$, which indicates that $\sigma_* / M \gtrsim
\mathcal{O}(10^{-1})$.  By using above equations, we can rewrite the
ratio in Eq.~\eqref{eq:ratio} as
\begin{equation}
\label{eq:ratio2}
\frac{P_{\zeta_\sigma}}{P_{\zeta_{\rm inf}}} 
\simeq 
3 \times 10^{14}  \epsilon \left( {100 \over f_{\rm NL} }\right)
\left( { 10^{10}\,\mbox{GeV} \over M }\right)^2. 
\end{equation}
Hence as far as $\epsilon$ is not extremely small, the fluctuations
from $\sigma$ in this mechanism can be responsible for cosmic
density perturbations today.

When the fluctuations from $\sigma$ are fully responsible for the
scalar primordial fluctuations, the tensor-to-scalar ratio is
\begin{eqnarray}
r = 8 \left( {\sigma_* \over M} \right)^{-4} \sigma_*^2 \simeq
 10^{-15} \times \left( \frac{f_{\rm NL}}{100} \right) 
\left( \frac{M}{10^{10} {\rm GeV}} \right)^2 ~.
\end{eqnarray}
Thus, as other mechanisms of generating primordial fluctuations such
as the curvaton and modulated reheating models, the
tensor-to-scalar ratio is generally very small in this mechanism.

Here we make some comments on $g_{\rm NL}$. With the form of $g$
assumed in Eq.~\eqref{eq:form_g} and when $g^{\prime\prime\prime}$ is
negligible, we have the relation between $f_{\rm NL}$ and $g_{\rm NL}$:
\begin{eqnarray}
g_{\rm NL} = -\frac{10}{3} f_{\rm NL} - \frac{50}{27}~.
\end{eqnarray}
Thus in this model, the sizes of $f_{\rm NL}$ and $g_{\rm NL}$ are
almost the same. 

In the above discussion, we have neglected the evolution of the light
field $\sigma$ after the horizon crossing until the end of thermal
inflation.  This assumption is valid when $m_{\sigma} \ll H_{\rm th}$
with $H_{\rm th}$ being the Hubble parameter during the thermal
inflation.  As we discussed in Section~\ref{sec:thermal}, 
$H_{\rm th}$ can be estimated as  $H_{\rm th} = V_0^{1/2}/ M_{\rm
  Pl} \simeq 10^{3}\,{\rm eV}$ and hence $\sigma$ has to be  light
as $m_\sigma \ll 10^3\,{\rm eV}$.  Moreover, $\sigma$ gets the
effective mass from the coupling $\sim g(\sigma) \phi^2 T^2$.  Thus
such an effective mass should be also smaller than the Hubble parameter
during the thermal inflation.  In our scenario, however, this effective
mass coming from the coupling can be negligible until the end of
thermal inflation because the flaton field is trapped at $\phi = 0$
during thermal inflation phase.

\section{Primordial fluctuations in scenarios with thermal inflation and 
models of inflation}
\label{sec:model}

Now in this section, we discuss what inflation models can be
compatible with the thermal inflation in the light of current
cosmological observations.  As mentioned in the introduction, when a
mini-inflation occurs after the primordial inflation, the time when the
fluctuations of the reference scale in the present observations exit
the horizon during inflation is shifted toward the end of inflation.
This, in most cases, means that the spectral index is more tilted.
Thus even if an inflation model consistent with current observations
in the standard case can be excluded when the thermal inflation
occurs.  However, this situation can be relaxed by adopting the
inhomogeneous end of thermal inflation discussed in
Section~\ref{sec:nonG} since fluctuations from the mechanism are
generally more scale-invariant as seen from Eq.~\eqref{eq:ns_2}.  We
investigate this issue by working out explicitly some inflation models
and discuss in what cases they are consistent with
current observations.

\subsection{The number of $e$-folds}

To predict the quantities such as the spectral index and
tensor-to-scalar ratio, we need to specify the number of $e$-folds
when the reference scale exits the horizon, at which above mentioned
quantities are measured.  The number of $e$-folds at the time when 
fluctuations of the scale $k$ exit  the
horizon  can be given by
\begin{eqnarray}
N_{\rm inf} = 50 - \ln \left( \frac{k}{a_0 H_0} \right) 
- \frac{2}{3} \ln \left( {10^{15} {\rm GeV} \over V_{\rm inf}^{1/4}}\right) 
- {1 \over 3}\ln \left( \frac{10^{6}\,\mbox{GeV}}{\rho_{\rm reh}^{1/4}}\right)~,
\label{eq:totefold}
\end{eqnarray}
where $V_{\rm inf}^{1/4}$ denotes the potential energy of the inflaton
during the inflation and $\rho_{\rm reh}$ denotes the energy density
of radiation at the reheating after the primordial inflation.  Here we
have assumed that the inflaton oscillates around a minimum of a
quadratic potential for some time after the inflation ends.

When the thermal inflation occurs, $N_{\rm inf}$ given above
would be reduced by $\Delta N_{\rm th}$ of
Eq.~(\ref{eq:efold_inc}).  Thus, in this case, the number of the
$e$-folds is modified as
\begin{eqnarray}
N_{\rm inf} &\! \simeq \!&  38 - \ln \left( \frac{k}{a_0 H_0} \right) 
 - \frac{2}{3} \ln \left( {10^{15} {\rm GeV} \over V_{\rm inf}^{1/4}}\right) 
- {1 \over 3}\ln \left( \frac{10^{6}\,\mbox{GeV}}{\rho_{\rm reh}^{1/4}}\right)
  \nonumber \\
&&  + \frac{1}{3}\ln\left( \frac{m}{ 10^2 {\rm GeV}}\right)
  + \frac{1}{3}\ln\left( \frac{T_r}{  {\rm GeV}}\right) 
  - \frac{2}{3}\ln\left( \frac{v_{\rm flaton}}{10^{10} {\rm GeV}}\right).
\label{eq:efold_thermal}
\end{eqnarray}
Since the spectral index and the tensor-to-scalar ratio depend 
on the number of the $e$-folds,
some models with  the reduced $N_{\rm inf}$ due to the thermal inflation 
may be excluded by current cosmological observations.

In the following, we consider some representative models of inflation
such as chaotic inflation, new inflation and hybrid inflation models.
First we focus on how the spectral index and tensor-to-scalar ratio
are modified in the presence of the thermal inflation when primordial
fluctuations are generated from the inflaton and discuss the
compatibility with observational constraints.  Then we investigate the
case where primordial fluctuations are produced by the
inhomogeneous end of thermal inflation, which is discussed in
Section~\ref{sec:nonG}, and see that some excluded inflation models can be
relaxed by using this mechanism.

\subsection{Chaotic inflation}

\begin{figure}[htbp]
  \begin{center}
    \resizebox{160mm}{!}{
    \includegraphics{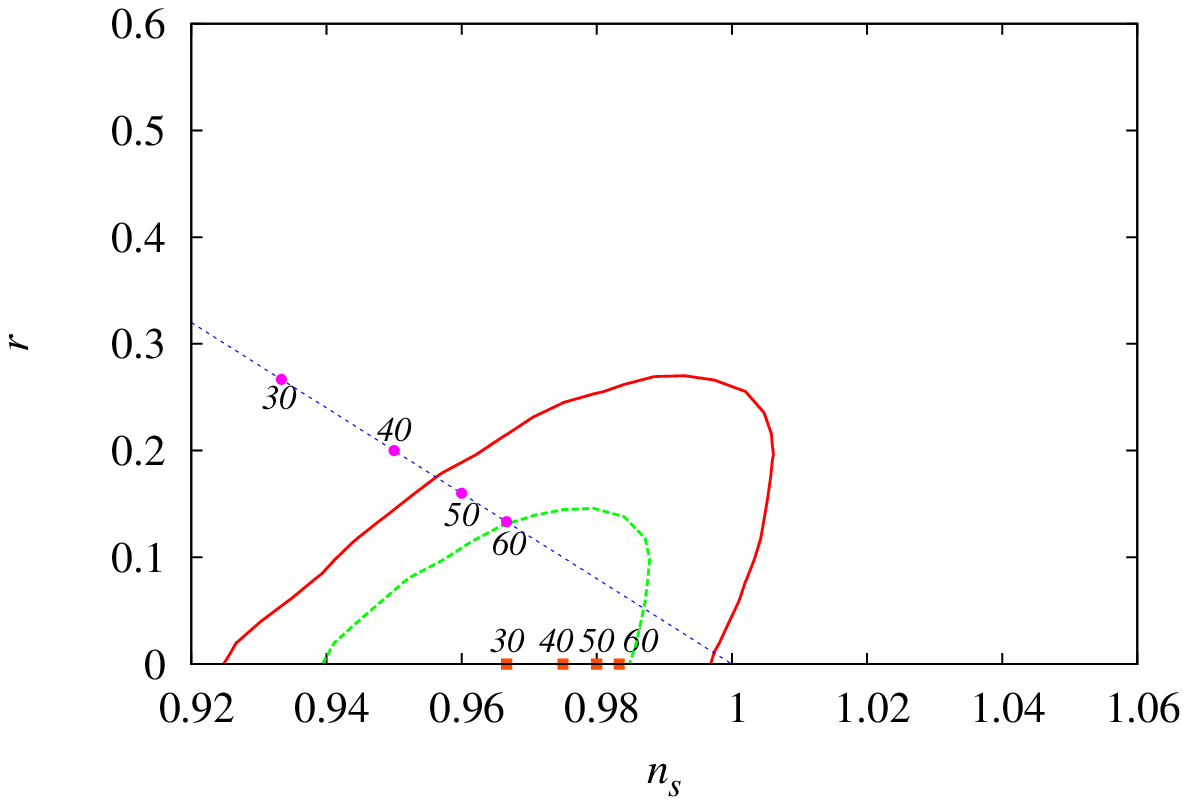}
    \includegraphics{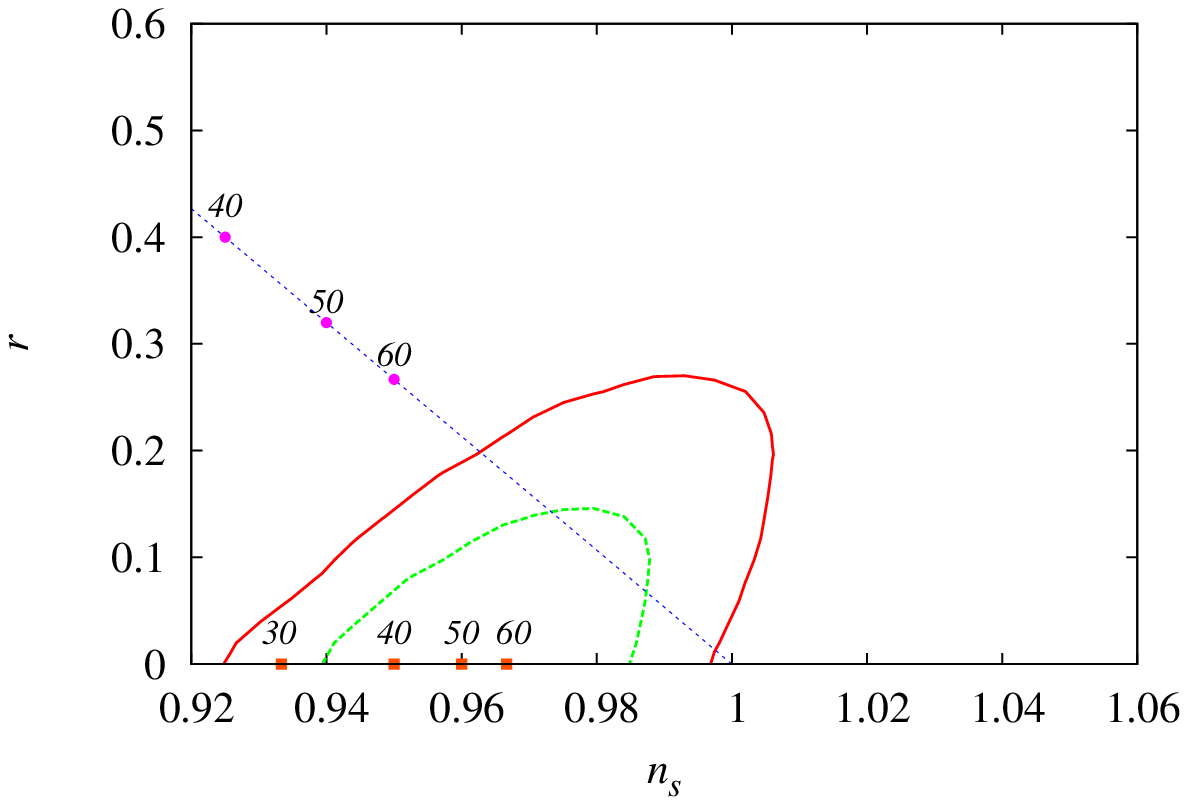}
    }
  \end{center}
  \caption{Contours indicate $68~\%$ (green dashed line) and $95~\%$
    (red solid line) C.L. allowed regions derived from WMAP+BAO+SN
    \cite{Komatsu:2008hk}.  The blue dotted line corresponds to the
    predictions for the quadratic (left) and the quartic (right)
    chaotic models. The small purple circles and red squares represent
    the values of $n_s$ and $r$ from the standard inflation and inhomogeneous 
    end of thermal inflation, respectively, for $N_{\rm inf} = 30, 40, 50$ and $60$.}
  \label{fig:wmap_bao_sn.eps}
\end{figure}

Chaotic inflation models have the potential of the following form:
\begin{equation}
V (\phi) = \lambda M_{\rm pl}^4 \left( \frac{\phi}{M_{\rm pl}} \right)^n,
\end{equation}
where $\phi$ is the inflaton field and $n$ is assumed to be even
integer to have a minimum at the origin.  The spectral index and the
tensor-to-scalar ratio in chaotic inflation models are respectively
given by, in terms of $N_{\rm inf}$,
\begin{eqnarray}
n_s - 1 = - {n + 2 \over 2 N_{\rm inf}}~,~r = { 4n \over N_{\rm inf} }~.
\end{eqnarray}
From the above expressions we can find that as $N_{\rm inf}$ becomes
smaller, $n_s$ is more red-tilted and $r$ becomes larger.  In
Fig.~\ref{fig:wmap_bao_sn.eps}, contours are shown for $68\,\%$ (green
dashed line) and $95\,\%$ (red solid line) C.L. allowed region derived
from WMAP+BAO+SN. The blue dotted line corresponds to the predictions
of chaotic inflation models on the $n_s$--$r$ plane for $n=2$ (left)
and $n=4$ (right) cases, respectively.  The values of $n_s$ and
$r$ in the standard inflation case are also shown with small circles
for $N_{\rm inf} = 30,40,50$ and $60$.  When $N_{\rm inf} \lesssim 45$, we
find that the quadratic model is excluded at $95\,\%$ C.L.
For the case of $n=4$, as is well-known, the
model is ruled out even if $N_{\rm inf}=60$.  However, if fluctuations
come from the inhomogeneous end of thermal inflation, the situation
drastically alters. As discussed in the previous section, the spectral
index is given by $n_s -1 = - 2\epsilon$ and the tensor-to-scalar
ratio would be negligibly small.  In Fig.~\ref{fig:wmap_bao_sn.eps},
we also plot the predictions of $n_s$ and $r$ from the inhomogeneous
end of thermal inflation with small squares.  Although the quadratic
chaotic inflation model with $N_{\rm inf} \lesssim 45$ are excluded by
current observations when the inflaton generates primordial
fluctuations, such a model becomes viable in the present mechanism
even if $N_{\rm inf}$ is small as 30.  More interestingly, the quartic
inflation model, which is excluded with its original form, can be viable
in the present mechanism as shown in the right panel of
Fig.~\ref{fig:wmap_bao_sn.eps}\footnote{
  For a similar argument in the framework of the mixed inflaton and
  curvaton or modulated reheating scenario, see \cite{mixed}.
}.

\subsection{new inflation}

For the new inflation model, we consider the following potential:
\begin{eqnarray}
V = V_0 \left[ 1 - 2 \left({\phi \over v}\right)^n + \left( {\phi \over v} \right)^{2n}\right]~,
\end{eqnarray}
where $v$ denotes the VEV of the inflaton.  We assume that $\phi / v
\ll 1$ during inflation.  Then, the slow-roll parameters are given by
\begin{eqnarray}
\epsilon = 2n^2 \left( {\phi_* \over v} \right)^{2(n-1)} \left( {M_{\rm pl} \over v} \right)^2~,
~ \eta = -2n(n-1)  \left( {\phi_* \over v} \right)^{n-2} \left( {M_{\rm pl} \over v} \right)^2~.
\end{eqnarray}
From these expressions, we find that $\epsilon / \eta = (\phi_* / v)^n
\ll 1$. Hence the tensor-to-scalar ratio is very small and the
spectral index is given by
\begin{eqnarray}
n_s- 1 \simeq 2 \eta~.
\end{eqnarray}
The $e$-folding number and the value of the inflaton field at the time
of horizon crossing are related by

\begin{eqnarray}
N_{\rm inf} = \left\{ \begin{array}{ll}
\displaystyle\frac{1}{4}\left(\frac{v}{M_{\rm pl}}\right)^2 \ln \left( \displaystyle\frac{\phi_e}{\phi_* }\right) \simeq 
\displaystyle\frac{1}{\eta} \left( \displaystyle\frac{\phi_e}{\phi_* }\right)& ~~(n=2) \\
\notag \\
\displaystyle\frac{1}{2n(n-2)}\left(\frac{v}{M_{\rm pl}}\right)^2 \left( \displaystyle\frac{\phi_*}{v} \right)^{-n+2} = - 
\left( \displaystyle\frac{n-1}{n-2}\right) {1 \over \eta} & ~~(n \ge 3) \\
\end{array} \right.
\end{eqnarray}
From these equations, the spectral index is given by, in terms of
$N_{\rm inf}$,
\begin{eqnarray}
n_s - 1 = \left\{ \begin{array}{ll}
 - \displaystyle\frac{2}{N_{\rm inf}} \left( \displaystyle\frac{\phi_e }{\phi_* }\right)& (n=2) \\
 \notag \\
 - \left( \displaystyle\frac{n-1}{n-2}\right) \displaystyle\frac{2}{N_{\rm inf}} & (n \ge 3) \\
\end{array} \right.
\end{eqnarray}
From the above expression, we find that when the curvature
perturbation is generated from the inflaton, $\left| n_s- 1 \right|$
becomes larger as $N_{\rm inf}$ becomes smaller.  For examples, when
$n=3$, $n_s=0.933,0.92$ and $0.9$ for the cases with $N_{\rm
  inf}=60,50$ and $40$, respectively.  From WMAP 5yr data, the limit
on $n_s$ is obtained as $ 0.925 < n_s < 0.997$ ($95\,\%$ C.L.) for
negligible tensor-to-scalar ratio \cite{Komatsu:2008hk}.  Hence, the
models with $N_{\rm inf} = 50$ and $40$ lie out of the $95\,\%$ C.L. allowed
region.

Now we consider the primordial fluctuations from the inhomogeneous end
of thermal inflation. As discussed in the previous section, the
condition $P_{\zeta_\sigma} / P_{\zeta_{\rm inf}} \gg 1 $ should be
satisfied when the fluctuations from the mechanism dominate over those
from the inflaton. However, this requires
$
\epsilon (\sigma_\ast / M)^2 (M_{\rm pl} / M)^2 \gg 1,
$
thus when $\epsilon$ is very small as in the case of the new
inflation, the above condition is difficult to be satisfied.  Even if
the fluctuations from the inflaton give a negligible contribution and
those from the inhomogeneous end of thermal inflation dominate, its
power spectrum becomes fairly scale-invariant since the spectral index
is given by $n_s -1 = -2 \epsilon \ll O(0.01)$.  Since the
tensor-to-scalar ratio is also very small, such fluctuations are also
excluded.  Thus, the new inflation may not be well suited to the
framework of the thermal inflation.

\subsection{Hybrid inflation}

Next we consider hybrid inflation. Although there are some variations
for the potential for the hybrid inflation, we discuss some
representative ones here.  First we consider the potential of the
following form:
\begin{equation}
\label{eq:V_hybrid1}
V(\phi) = V_0 + \frac{1}{2} m_\phi^2 \phi^2.
\end{equation}
When 
$V_0 \gg  m_\phi^2 \phi^2,$
the slow-roll parameter can be written as 
\begin{equation}
\epsilon \simeq \frac{M_{\rm pl}^2}{2}\frac{m_\phi^4 \phi^2}{V_0^2},
~~~~~~~~~~~
\eta \simeq \frac{m^2_\phi M_{\rm pl}^2}{V_0}.
\end{equation}
Since the condition 
$V_0 \gg  m_\phi^2 \phi^2$
implies that 
$ \epsilon \ll \eta,$ the spectral index of this model can be written
as $n_s -1 = 2\eta$, which is blue-tilted since $\eta$ is positive
here.  The smallness of $\epsilon$ also indicates that the
tensor-to-scalar ratio is also negligible in this model,
and in such a case the blue-tilted spectrum conflicts with the
observations (see Fig.~\ref{fig:wmap_bao_sn.eps}).
Thus, this
type of hybrid inflation is excluded by current observations in its
original form.  As in the case of the new inflation, since $\epsilon$
is very small in this model as well, it could be difficult to have the
case where fluctuations from the inhomogeneous end of thermal
inflation dominate over those from the inflaton. Furthermore 
even if the fluctuations from the
mechanism dominates over those from the inflaton, the spectral index
is $n_s \simeq 1$, thus this again contradicts with current
observations.

Another possible realization of the hybrid inflation is so-called
mutated hybrid inflation \cite{Stewart:1994pt}, in which the potential
can be written as
\begin{equation}
V(\phi) = V_0 - \mu M_{\rm pl}^4  \left(\frac{M_{\rm pl}}{\phi} \right)^n,
\end{equation}
where $\mu$ is a constant. The potential of this kind can appear, for
example, smooth hybrid inflation \cite{Lazarides:1995vr} for $n=4$.
With this potential, the slow-roll parameters are
\begin{equation}
\epsilon \simeq 
\frac{\mu^2 n^2}{2} \left( \frac{M_{\rm pl}^4}{V_0}\right)^2
 \left( \frac{M_{\rm pl}}{\phi} \right)^{2(n+1)},
~~~~~~
\eta \simeq - \mu n(n+1) \left( \frac{M_{\rm pl}^4}{V_0}\right)
 \left( \frac{M_{\rm pl}}{\phi} \right)^{n+2},
\end{equation}
where we assumed that 
$
V_0 \gg \mu M_{\rm pl}^4  \left( \frac{M_{\rm pl}}{\phi} \right)^n.
$
This condition also implies that $\epsilon \ll |\eta|$, thus the
spectral index can be given by
\begin{equation}
n_s -1 = - \frac{2(n+1)}{(n+2)N_{\rm inf}}.
\end{equation}
When $n=2$, the spectral index is $n_s = 0.95, 0.962, 0.97$ and
$0.975$ for $N_{\rm inf} = 30, 40, 50$ and $60$, respectively.  Thus
even if the number of $e$-folds is reduced due to the thermal
inflation, this model is not excluded. In fact, the central values of
the allowed range for the spectral index from WMAP5 analysis is $n_s
\sim 0.96$, thus a slight reduction of $N_{\rm inf}$ may be favored.
It should be noticed that the slow-roll parameter $\epsilon$ in this
model could also be very small, the fluctuations from the
inhomogeneous end of thermal inflation may not give a dominant
contribution for the curvature fluctuations. Even if  fluctuations from 
the inhomogeneous end of thermal inflation are
 the main component of $\zeta$, the spectral index is again too
scale-invariant to be consistent with observations, which was also the
case for the new inflation and hybrid inflation with the potential of
Eq.~\eqref{eq:V_hybrid1}.  The mutated hybrid inflation discussed here
can be compatible to cosmological observations in the framework of
thermal inflation even in its original form.

\section{Summary}\label{sec:summary}

In this paper, we have presented a mechanism of generating primordial
curvature perturbation at the end of thermal inflation by considering
the case where the coupling of a flaton field with the fields in
thermal bath can fluctuate.  We also show that there is a possibility
of generating large non-Gaussianity in this scenario.

We have also investigated the constraint on inflation models in the
case where the thermal inflation is realized after the primordial
inflation.  When such a mini-inflation occurs, the time when the
fluctuations of the reference scale in the present observations exit
the horizon during inflation is shifted toward the end of inflation.
This, in most cases, means that the spectral index is more tilted.
Thus even if an inflation model is consistent with current
observations in the standard case, some models can be excluded when
the thermal inflation occurs. We have discussed that this situation
can be relaxed by adopting the inhomogeneous end of thermal inflation
since fluctuations from the mechanism are generally more
scale-invariant as seen from Eq.~\eqref{eq:ns_2}.  We have explicitly
worked on some simple inflation models, such as chaotic inflation, new
inflation and hybrid inflation. Then we find that some excluded models
can be relaxed by using this mechanism proposed in this paper.

As a final comment,  we mention  baryogenesis in a scenario with 
the thermal inflation. Since the  thermal inflation dilutes the preexisting 
baryon asymmetry, one needs to generate the baryon number 
after thermal inflation. Some authors have
discussed Affleck-Dine baryogenesis after thermal inflation~\cite{Choi:2009qd}. 
It may  also be interesting to investigate observational signature in such
model with the inhomogeneous end of thermal inflation scenario, for
example, baryon isocurvature fluctuations. This would be a subject of
the future study.

\section*{Acknowledgments}

This work is supported by Grant-in-Aid for Scientific research from
the Ministry of Education, Science, Sports, and Culture, Japan,
No. 14102004 (M.K.) and No. 19740145 (T.T.), and also by World Premier
International Research Center Initiative, MEXT, Japan.  S.Y. is
supported in part by Grant-in-Aid for Scientific Research on Priority
Areas No. 467 ``Probing the Dark Energy through an Extremely Wide and
Deep Survey with Subaru Telescope''.  He also acknowledges the support
from the Grant-in-Aid for the Global COE Program ``Quest for
Fundamental Principles in the Universe: from Particles to the Solar
System and the Cosmos'' from MEXT, Japan.


\begin{thebibliography}{99}


\bibitem{Yamamoto:1985rd}
  K.~Yamamoto,
  Phys.\ Lett.\  B {\bf 168}, 341 (1986).


\bibitem{Binetruy:1986ss}
  P.~Binetruy and M.~K.~Gaillard,
  Phys.\ Rev.\  D {\bf 34}, 3069 (1986).

\bibitem{Lazarides:1992gg}
  G.~Lazarides, C.~Panagiotakopoulos and Q.~Shafi,
  Phys.\ Rev.\ Lett.\  {\bf 56}, 557 (1986);
  G.~Lazarides, C.~Panagiotakopoulos and Q.~Shafi,
  Phys.\ Rev.\ Lett.\  {\bf 58}, 1707 (1987);
  G.~Lazarides and Q.~Shafi,
  Nucl.\ Phys.\  B {\bf 392}, 61 (1993).

\bibitem{Lyth:1995hj}
  D.~H.~Lyth and E.~D.~Stewart,
  Phys.\ Rev.\ Lett.\  {\bf 75}, 201 (1995)
  [arXiv:hep-ph/9502417].

\bibitem{Lyth:1995ka}
  D.~H.~Lyth and E.~D.~Stewart,
  Phys.\ Rev.\  D {\bf 53}, 1784 (1996)
  [arXiv:hep-ph/9510204].


\bibitem{Barreiro:1996dx}
  T.~Barreiro, E.~J.~Copeland, D.~H.~Lyth and T.~Prokopec,
  Phys.\ Rev.\  D {\bf 54}, 1379 (1996)
  [arXiv:hep-ph/9602263].



\bibitem{Asaka:1999xd}
  T.~Asaka and M.~Kawasaki,
  Phys.\ Rev.\  D {\bf 60}, 123509 (1999)
  [arXiv:hep-ph/9905467].


\bibitem{Coughlan:1983ci}
  G.~D.~Coughlan, W.~Fischler, E.~W.~Kolb, S.~Raby and G.~G.~Ross,
  Phys.\ Lett.\  B {\bf 131}, 59 (1983).


\bibitem{Banks:1993en}
  T.~Banks, D.~B.~Kaplan and A.~E.~Nelson,
  Phys.\ Rev.\  D {\bf 49}, 779 (1994)
  [arXiv:hep-ph/9308292].
  
\bibitem{deCarlos:1993jw}
  B.~de Carlos, J.~A.~Casas, F.~Quevedo and E.~Roulet,
  Phys.\ Lett.\  B {\bf 318}, 447 (1993)
  [arXiv:hep-ph/9308325].
  

\bibitem{Komatsu:2008hk}
  E.~Komatsu {\it et al.}  [WMAP Collaboration],
  Astrophys.\ J.\ Suppl.\  {\bf 180}, 330 (2009)
  [arXiv:0803.0547 [astro-ph]].


\bibitem{Enqvist:2001zp}
K.~Enqvist and M.~S.~Sloth,
Nucl.\ Phys.\ B {\bf 626}, 395 (2002)
[arXiv:hep-ph/0109214].
\bibitem{Lyth:2001nq}
D.~H.~Lyth and D.~Wands,
Phys.\ Lett.\ B {\bf 524}, 5 (2002)
[arXiv:hep-ph/0110002].
\bibitem{Moroi:2001ct}
T.~Moroi and T.~Takahashi,
Phys.\ Lett.\ B {\bf 522}, 215 (2001)
[Erratum-ibid.\ B {\bf 539}, 303 (2002)]
[arXiv:hep-ph/0110096].


\bibitem{Dvali:2003em}
  G.~Dvali, A.~Gruzinov and M.~Zaldarriaga,
  Phys.\ Rev.\  D {\bf 69}, 023505 (2004)
  [arXiv:astro-ph/0303591].
\bibitem{Kofman:2003nx}
  L.~Kofman,
  arXiv:astro-ph/0303614.



\bibitem{Kolb:2004jm}
  E.~W.~Kolb, A.~Riotto and A.~Vallinotto,
  Phys.\ Rev.\  D {\bf 71}, 043513 (2005)
  [arXiv:astro-ph/0410546].

\bibitem{Lyth:2005qk}
  D.~H.~Lyth,
  JCAP {\bf 0511}, 006 (2005)
  [arXiv:astro-ph/0510443].
  
\bibitem{Alabidi:2006wa}
  L.~Alabidi and D.~Lyth,
  JCAP {\bf 0608}, 006 (2006)
  [arXiv:astro-ph/0604569].

\bibitem{Sasaki:2008uc}
  M.~Sasaki,
  Prog.\ Theor.\ Phys.\  {\bf 120}, 159 (2008)
  [arXiv:0805.0974 [astro-ph]].

\bibitem{Smith:2009jr}
  K.~M.~Smith, L.~Senatore and M.~Zaldarriaga,
  arXiv:0901.2572 [astro-ph].


\bibitem{nonGcurvaton}
  N.~Bartolo, S.~Matarrese and A.~Riotto,
  Phys.\ Rev.\  D {\bf 69}, 043503 (2004)
  [arXiv:hep-ph/0309033];
  K.~Enqvist and S.~Nurmi,
  JCAP {\bf 0510}, 013 (2005)
  [arXiv:astro-ph/0508573];
  K.~A.~Malik and D.~H.~Lyth,
  JCAP {\bf 0609}, 008 (2006)
  [arXiv:astro-ph/0604387];
  M.~Sasaki, J.~Valiviita and D.~Wands,
  Phys.\ Rev.\  D {\bf 74}, 103003 (2006)
  [arXiv:astro-ph/0607627];
  Q.~G.~Huang,
  arXiv:0801.0467 [hep-th];
  K.~Ichikawa, T.~Suyama, T.~Takahashi and M.~Yamaguchi,
  Phys.\ Rev.\  D {\bf 78}, 023513 (2008)
  [arXiv:0802.4138 [astro-ph]];
  K.~Enqvist and T.~Takahashi,
  JCAP {\bf 0809}, 012 (2008)
  [arXiv:0807.3069 [astro-ph]];
  Q.~G.~Huang,
  JCAP {\bf 0811}, 005 (2008)
  [arXiv:0808.1793 [hep-th]];
  Q.~G.~Huang and Y.~Wang,
  JCAP {\bf 0809}, 025 (2008)
  [arXiv:0808.1168 [hep-th]];
  T.~Moroi and T.~Takahashi,
  Phys.\ Lett.\  B {\bf 671}, 339 (2009)
  [arXiv:0810.0189 [hep-ph]];
M.~Kawasaki, K.~Nakayama and F.~Takahashi,
JCAP {\bf 0901} (2009) 026
[arXiv:0810.1585 [hep-ph]];
  P.~Chingangbam and Q.~G.~Huang,
  JCAP {\bf 0904}, 031 (2009)
  [arXiv:0902.2619 [astro-ph.CO]];
  K.~Enqvist and T.~Takahashi,
  arXiv:0909.5362 [astro-ph.CO].


\bibitem{nonGmodreheat}


  M.~Zaldarriaga,
  Phys.\ Rev.\  D {\bf 69}, 043508 (2004)
  [arXiv:astro-ph/0306006];
  T.~Suyama and M.~Yamaguchi,
  Phys.\ Rev.\  D {\bf 77}, 023505 (2008)
  [arXiv:0709.2545 [astro-ph]];
 K.~Ichikawa, T.~Suyama, T.~Takahashi and M.~Yamaguchi,
  Phys.\ Rev.\  D {\bf 78}, 063545 (2008)
  [arXiv:0807.3988 [astro-ph]];
  T.~Takahashi, M.~Yamaguchi and S.~Yokoyama,
  arXiv:0907.3052 [astro-ph.CO].


\bibitem{Matsuda:2009yt}
  T.~Matsuda,
  Class.\ Quant.\ Grav.\  {\bf 26}, 145011 (2009)
  [arXiv:0902.4283 [hep-ph]].


\bibitem{bib:deltaN}
  A.~A.~Starobinsky,
  JETP Lett.\  {\bf 42} (1985) 152
  [Pisma Zh.\ Eksp.\ Teor.\ Fiz.\  {\bf 42} (1985) 124];
M.~Sasaki and E.~D.~Stewart,
  Prog.\ Theor.\ Phys.\  {\bf 95}, 71 (1996).
  [arXiv:astro-ph/9507001]; 
  M.~Sasaki and T.~Tanaka,
  Prog.\ Theor.\ Phys.\  {\bf 99}, 763 (1998).
  [arXiv:gr-qc/9801017].


\bibitem{Lyth:2004gb}
  D.~H.~Lyth, K.~A.~Malik and M.~Sasaki,
  JCAP {\bf 0505}, 004 (2005)
  [arXiv:astro-ph/0411220].
\bibitem{Lyth:2005fi}
  D.~H.~Lyth and Y.~Rodriguez,
  Phys.\ Rev.\ Lett.\  {\bf 95}, 121302 (2005)
  [arXiv:astro-ph/0504045];
  L.~Alabidi and D.~H.~Lyth,
  JCAP {\bf 0605}, 016 (2006)
  [arXiv:astro-ph/0510441];
  C.~T.~Byrnes, M.~Sasaki and D.~Wands,
  Phys.\ Rev.\  D {\bf 74}, 123519 (2006)
  [arXiv:astro-ph/0611075].


 
\bibitem{mixed}
  D.~Langlois and F.~Vernizzi,
  Phys.\ Rev.\  D {\bf 70}, 063522 (2004)
  [arXiv:astro-ph/0403258];
  T.~Moroi, T.~Takahashi and Y.~Toyoda,
  Phys.\ Rev.\  D {\bf 72}, 023502 (2005)
  [arXiv:hep-ph/0501007];
  T.~Moroi and T.~Takahashi,
  Phys.\ Rev.\  D {\bf 72}, 023505 (2005)
  [arXiv:astro-ph/0505339];
   K.~Ichikawa, T.~Suyama, T.~Takahashi and M.~Yamaguchi, in Ref.~\cite{nonGcurvaton};
   K.~Ichikawa, T.~Suyama, T.~Takahashi and M.~Yamaguchi, in Ref.~\cite{nonGmodreheat}.
   

\bibitem{Stewart:1994pt}
  E.~D.~Stewart,
  Phys.\ Lett.\  B {\bf 345}, 414 (1995)
  [arXiv:astro-ph/9407040].

\bibitem{Lazarides:1995vr}
  G.~Lazarides and C.~Panagiotakopoulos,
  Phys.\ Rev.\  D {\bf 52}, R559 (1995)
  [arXiv:hep-ph/9506325].


\bibitem{Choi:2009qd}

  E.~D.~Stewart, M.~Kawasaki and T.~Yanagida,
  Phys.\ Rev.\  D {\bf 54}, 6032 (1996)
  [arXiv:hep-ph/9603324];
  D.~h.~Jeong, K.~Kadota, W.~I.~Park and E.~D.~Stewart,
  JHEP {\bf 0411}, 046 (2004)
  [arXiv:hep-ph/0406136];
  M.~Kawasaki and K.~Nakayama,
  Phys.\ Rev.\  D {\bf 74}, 123508 (2006)
  [arXiv:hep-ph/0608335];
  G.~N.~Felder, H.~Kim, W.~I.~Park and E.~D.~Stewart,
  JCAP {\bf 0706}, 005 (2007)
  [arXiv:hep-ph/0703275];
  S.~Kim, W.~I.~Park and E.~D.~Stewart,
  JHEP {\bf 0901}, 015 (2009)
  [arXiv:0807.3607 [hep-ph]];
  K.~Choi, K.~S.~Jeong, W.~I.~Park and C.~S.~Shin,
  arXiv:0908.2154 [hep-ph].

\end{thebibliography}
\end{document}